\documentclass[aps,pra,reprint,showpacs,groupedaddress]{revtex4-1}

\usepackage[utf8]{inputenc}
\usepackage{amsmath,amssymb,amsfonts}
\usepackage{graphicx}
\usepackage{psfrag}

\usepackage{color}

\usepackage{hyperref}

\psfrag{1}{$r_0$}
\psfrag{3}{$\Omega$}
\psfrag{4}{}
\psfrag{2}{$2R$}
\psfrag{m}[c]{Excitation quantum number $m$}
\psfrag{energy}[c]{Excitation energy $\hbar\omega$~[units of $\hbar\omega_r]$}
\psfrag{r0}[c]{Ring radius $r_0$~[units of $a_r]$}
\psfrag{omega_c}[c]{$\Omega_c^{-1} $~[units of $\omega_r^{-1}]$}
\psfrag{omega2}[tc]{$\Omega_c$~[units of $\omega_r]$}
\psfrag{mu}[c]{$\mu$~[u. of $\hbar\omega_r]$}
\psfrag{m_l}[c]{Excitation quantum number $m$}
\psfrag{energy_l}[c]{Excitation energy $\hbar\omega$~[units of $\hbar\omega_r]$}
\psfrag{m_l2}[c]{$m$}
\psfrag{energy_l2}[c]{$\hbar\omega$~[u. of $\hbar\omega_r$]}
\psfrag{density}{Density~[a.u.]}
\psfrag{radius}{radius~[units of $a_r]$}
\psfrag{a}[c]{(a)}
\psfrag{b}[c]{(b)}
\psfrag{l0}[c]{$\ell=0$}
\psfrag{l15}[c]{$\ell=15$}
\psfrag{c}[c]{\textcolor{white}{(c)}}
\psfrag{d}[c]{\textcolor{white}{(d)}}
\psfrag{l-m}[c]{Initial state circulation $\ell$}
\psfrag{omega-l}[c]{Critical angular velocity $\Omega_c$~[units of $\omega_r]$}
\psfrag{l1}{$\ell_1$}
\psfrag{l2}{$\ell_2$}
\psfrag{l-m2}[c]{radius~[units of $a_r]$}
\psfrag{omega-l2}[c]{Slope~[units of $\omega_r]$}
\psfrag{in}[c]{inner}
\psfrag{out}[c]{outer}

\newcommand{\abs}[1]{\left|#1\right|}

\newcommand{\re}[1]{\textrm{Re}\left[#1\right]}

\begin{document}


\title{Critical rotation of an annular superfluid Bose gas}


\author{R. Dubessy}
\author{T. Liennard}
\author{P. Pedri}
\author{H. Perrin}
\email[]{helene.perrin@univ-paris13.fr}
\affiliation{Laboratoire de physique des lasers, CNRS and Université
Paris 13, 99 avenue J.-B. Clément, F-93430 Villetaneuse}


\date{\today}

\begin{abstract}
We analyze the excitation spectrum of a superfluid Bose-Einstein
condensate rotating in a ring trap.
We identify two important branches of the spectrum related to outer
and inner edge surface modes that lead to the instability of the
superfluid.
Depending on the initial circulation of the annular condensate, either
the outer or the inner modes become first unstable.
This instability is crucially related to the superfluid nature of the
rotating gas.
In particular we point out the existence of a maximal circulation
above which the superflow decays spontaneously, which cannot be
explained by invoking the average speed of sound.
\end{abstract}


\maketitle

After the pioneering work on persistent flow in
helium~\cite{Hall1957}, recent experimental success at producing
circulating superfluid flow of Bose gases in annular
traps~\cite{Ryu2007,Ramanathan2011,Moulder2012} has focused interest
on the issue of dissipation of this macroscopic quantum state.
In a superfluid this question is crucially related to the existence of
a critical velocity $v_c$ above which excitations are generated.
The critical velocity is determined by the Landau criterion~\cite{Landau1941}.
Dissipation occurs for a fluid velocity larger than $v_c$.
Symmetrically a defect moving above $v_c$ generates excitations in a
superfluid at rest.
This has been evidenced experimentally in trapped Bose gases~\cite{Raman1999}.

In a homogeneous weakly interacting Bose gas $v_c$ is equal to the
speed of sound~\cite{Leggett1999}.
This is no longer true if the system is inhomogeneous.
For example, in an infinite cylindrically symmetric tube with
transverse harmonic confinement, the critical velocity is lower than
the speed of sound~\cite{Fedichev2001}.
In such a geometry, the first modes excited at the critical velocity
have been shown to be surface modes~\cite{Anglin2001} propagating
along the tube and localized symmetrically all around the edge of the
cylinder.

\begin{figure}[t]
\includegraphics[width=5cm]{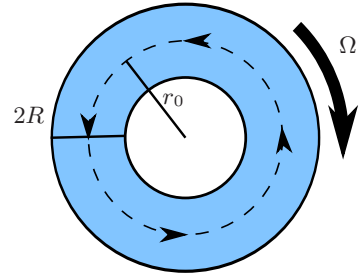}
\caption{\label{fig:ring} (Color online)
Sketch of the system: A Bose gas flowing with a circulation $\ell$
(dashed line) and a Thomas-Fermi width $2R$ is held in a ring trap of
radius $r_0$.
A defect rotating counterflow at angular velocity $\Omega$ above the
critical velocity will induce dissipation.
}
\end{figure}

In order to study superfluidity experimentally,
it is natural to bind this system and investigate the stability of a
persistent flow in a ring geometry.
A crucial difference between a tube and a ring is the
presence of a centrifugal force arising from the non-Galilean nature
of rotation.
Moreover, the curvature of the annulus makes the inner and the outer
surfaces of the fluid no longer equivalent~\cite{Donnelly1966} (see
Fig.~\ref{fig:ring}).

The ring geometry has recently attracted a lot of interest.
Many annular traps have been
proposed~\cite{Wright2000,Morizot2006,Lesanovsky2007} and
realized~\cite{Ryu2007,Heathcote2008,Ramanathan2011,Sherlock2011,Moulder2012}.
Studies of the superfluidity include the observation of a persistent
current~\cite{Ryu2007}, the effect of a weak
link~\cite{Ramanathan2011,Piazza2009,Schenke2011}, and the observation
of a stepwise dissipation of the circulation~\cite{Moulder2012}. The
ground state of the system in the presence of rotation has been determined
theoretically~\cite{Fetter1967,Cozzini2006,Aftalion2010}. Phase
fluctuations in a ring trap have also been
investigated~\cite{Mathey2010}.
However, the determination of the critical angular velocity in a ring
is still an open question and is highly relevant to recent
experiments~\cite{Ramanathan2011,Sherlock2011,Moulder2012}.

In this Rapid Communication we determine the critical angular velocity in the sense
of the Landau criterion for a Bose gas trapped in a ring. We compute
the Bogoliubov excitation spectrum both for an initially non-rotating
gas in the ground state, and for an initially circulating stationary
state.
We show that the critical velocity is governed by surface modes, like
in the case of an infinite tube.
However, we find that there are now two distinct nondegenerate
families of surface excitations propagating either at the inner or at
the outer edge.
The lowest of these two branches gives the critical angular velocity.
Our numerical calculations predict the existence of a maximal
circulation, above which the system becomes unstable, any static
perturbation giving rise to dissipation of the flow.
We give a simple interpretation of all these features by extending the
surface mode model~\cite{Anglin2001} to the ring geometry.
Our model is in good agreement with the numerical calculations even
for an initially circulating state.

We consider a condensate confined in a ring-shaped trap (see
Fig.~\ref{fig:ring}) described by the Gross-Pitaevskii equation at
zero temperature.
For simplicity, we reduce the problem to two dimensions in the
plane of the ring.
The two-dimensional (2D) ring still allows one to identify the inner and outer edges and
evidence their respective roles.
The trapping annular potential is written as a harmonic potential of
frequency $\omega_r$ centered at a radius $\rho_0$.
In the following, we use the associated scales for energy
($\hbar\omega_r$), time ($\omega_r^{-1}$) and length
$a_r=\sqrt{\hbar/(M\omega_r)}$, where $M$ is the atomic mass. The 2D
Gross-Pitaevskii equation reads
\begin{equation}
i\partial_t\psi=\left(-\frac{\Delta}{2}+\frac{1}{2}(r-r_0)^2+g\abs{\psi}^2\right)\psi,
\label{eqn:gpe2D}
\end{equation}
where $\psi=\psi(r,\theta,t)$ is normalized to unity,
$\Delta=\partial_r^2+\partial_r/r+\partial_\theta^2/r^2$ is the
Laplacian in polar coordinates $(r,\theta)$, $r_0=\rho_0/a_r$ is the
dimensionless ring radius, and $g$ is the dimensionless interaction
constant in two dimensions~\footnote{$g=N\sqrt{8\pi}\frac{a}{a_z}$ where $N$ is
the atom number, $a$ is the scattering length and $a_z$ the ground
state size along the strongly confined direction $z$, see for
example~\cite{Petrov2000}.}.

Using the rotational invariance of Eq.~\eqref{eqn:gpe2D}, we consider
solutions of the form:
\begin{eqnarray}
&&\psi(r,\theta,t)=e^{-i(\mu
t-\ell\theta)}\left[\psi_\ell(r)+\delta\psi_m^\ell(r,\theta,t)\right],\\
&&\textrm{where}\nonumber\\
&&\delta\psi_m^\ell(r,\theta,t)=u_m^\ell(r)e^{-i(\omega
t-m\theta)}+v_m^\ell(r)^*e^{i(\omega^* t-m\theta)}.
\end{eqnarray}
The stationary solution $\psi_\ell(r)$ is a state of circulation
$\ell$ and chemical potential $\mu$, which depends on $\ell$, and
$\delta\psi_m^\ell$ is a small perturbation, parametrized by $\ell$
and $m$.
$\psi_\ell(r)$ is not necessarily the ground state of the system but
can be realized experimentally using phase
imprinting~\cite{Ramanathan2011,Moulder2012}.
We label as $R=\sqrt{2\mu}$ the half width of the radial density
distribution in the Thomas-Fermi approximation.

Linearizing Eq.~\eqref{eqn:gpe2D}, we solve the Bogoliubov--de Gennes
equations for $u_m^\ell(r)$ and $v_m^\ell(r)$~\footnote{The initial
stationary state of circulation $\ell$ is found as the result of an
imaginary time propagation of a test Thomas-Fermi profile stopped when
the relative variation of the chemical potential reaches $10^{-12}$.
The Bogoliubov-de Gennes equations are diagonalized using a C++
implementation of the LAPACK library~\cite{Anderson1999}.}.
We get real frequencies with a dispersion relation
$\omega=\omega_\ell(m)$ for each initial circulation $\ell$.
The lowest branch of the spectrum allows us to compute the critical
angular velocity
$\Omega_c(\ell)=\min_{m}\left[\omega_\ell(m)/\abs{m}\right]$ for a
given circulation $\ell$.

\begin{figure}[t]
\includegraphics[width=\columnwidth]{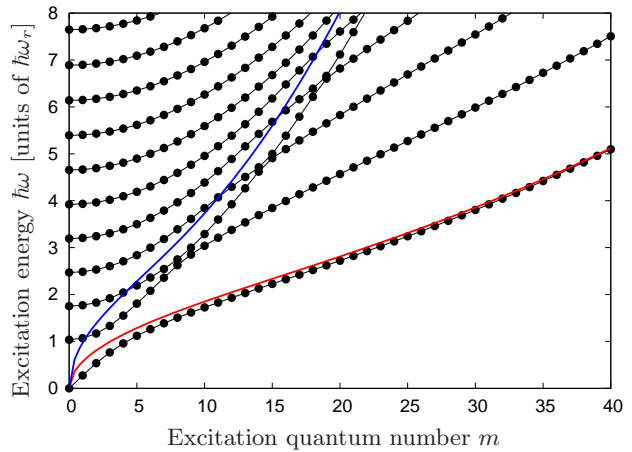}%
\caption{\label{fig:spectrum} (Color online) Excitation spectrum
obtained from Eq.~\eqref{eqn:gpe2D} (see text for details), with $\ell
= 0$, $r_0=12$ and $g=9000$.
Only the lowest eleven branches are shown.
The solid lines are a guide to the eye to distinguish between the
different branches.
A symmetric spectrum also exists for negative values of $m$.
Solid lines: relation dispersion in the surface mode model for the
inner (upper blue line) and outer (lower red line) modes.
}
\end{figure}

\begin{figure}[t]
\includegraphics[width=\columnwidth]{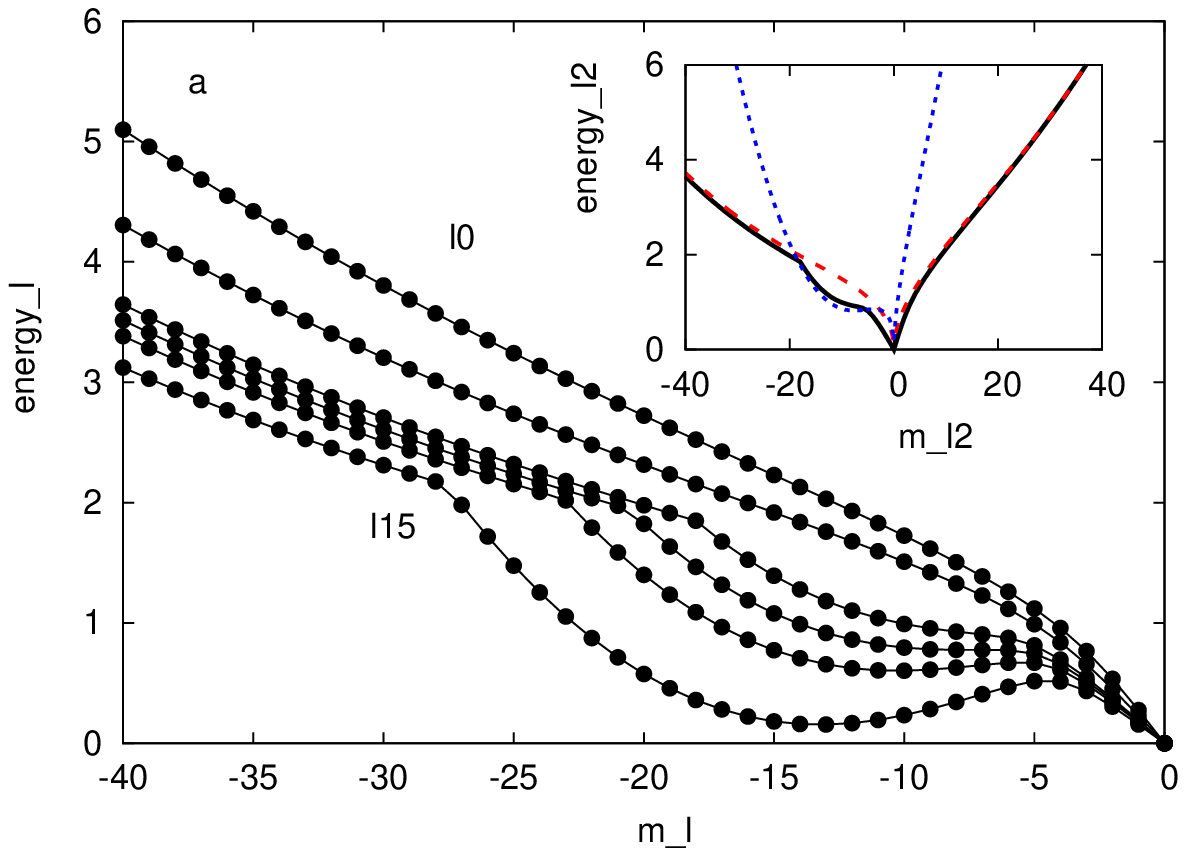}\\
\includegraphics[width=\columnwidth]{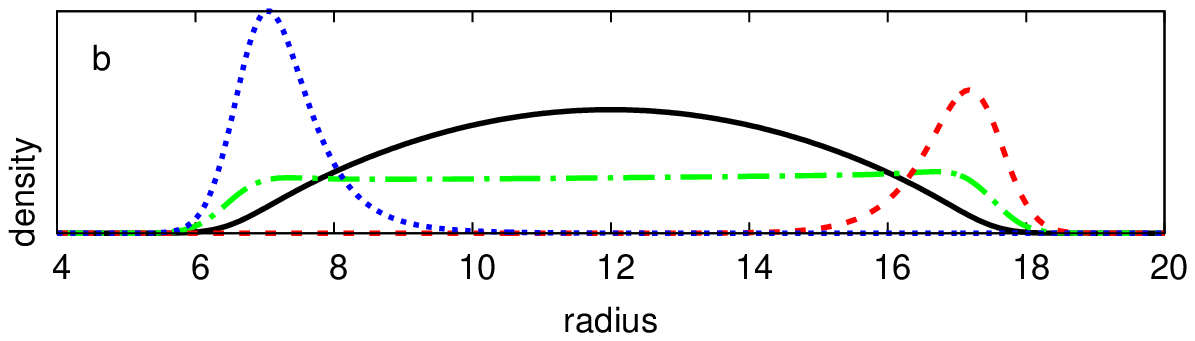}\\
\vspace{3mm}
\includegraphics[width=0.4\columnwidth]{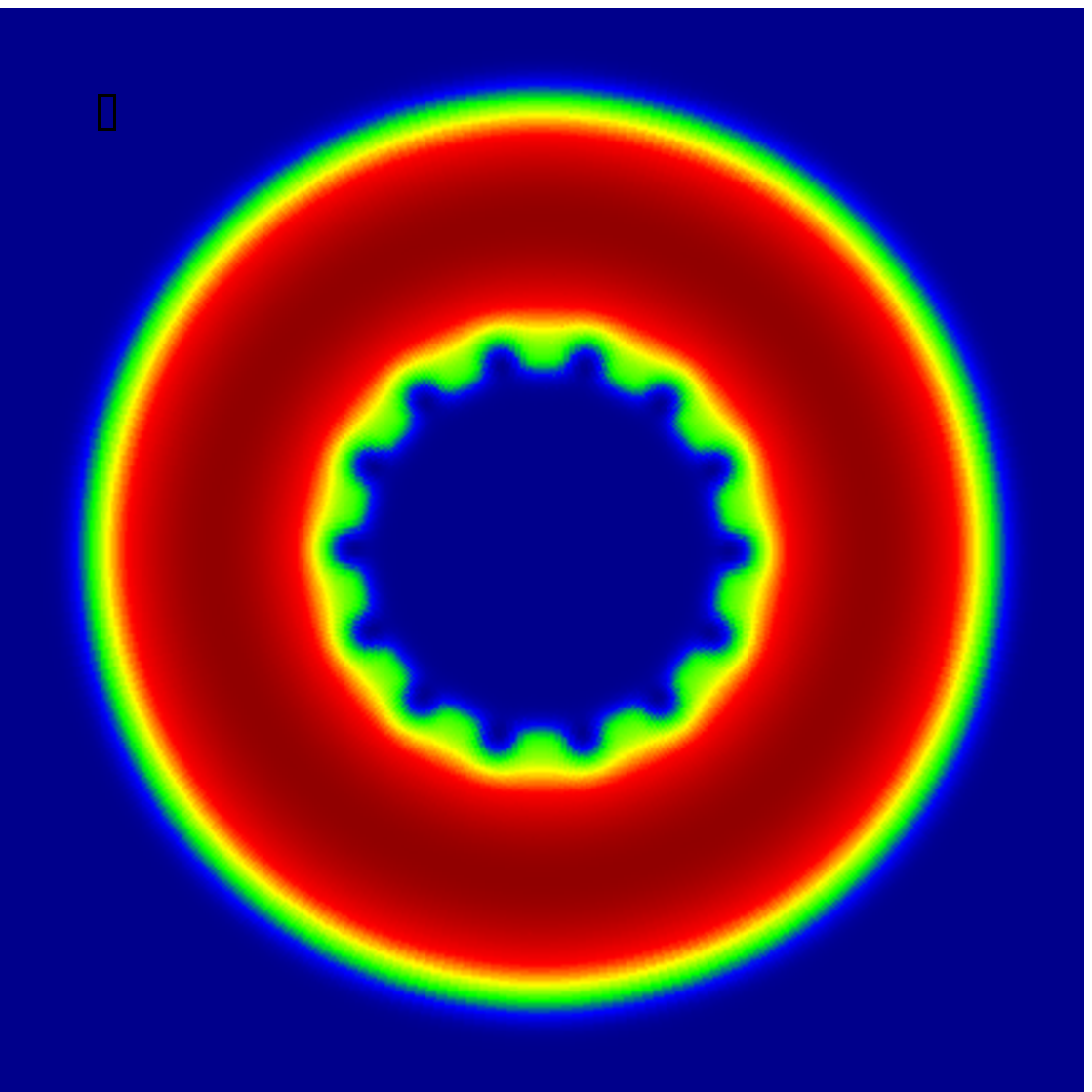}\hspace{0.7cm}
\includegraphics[width=0.4\columnwidth]{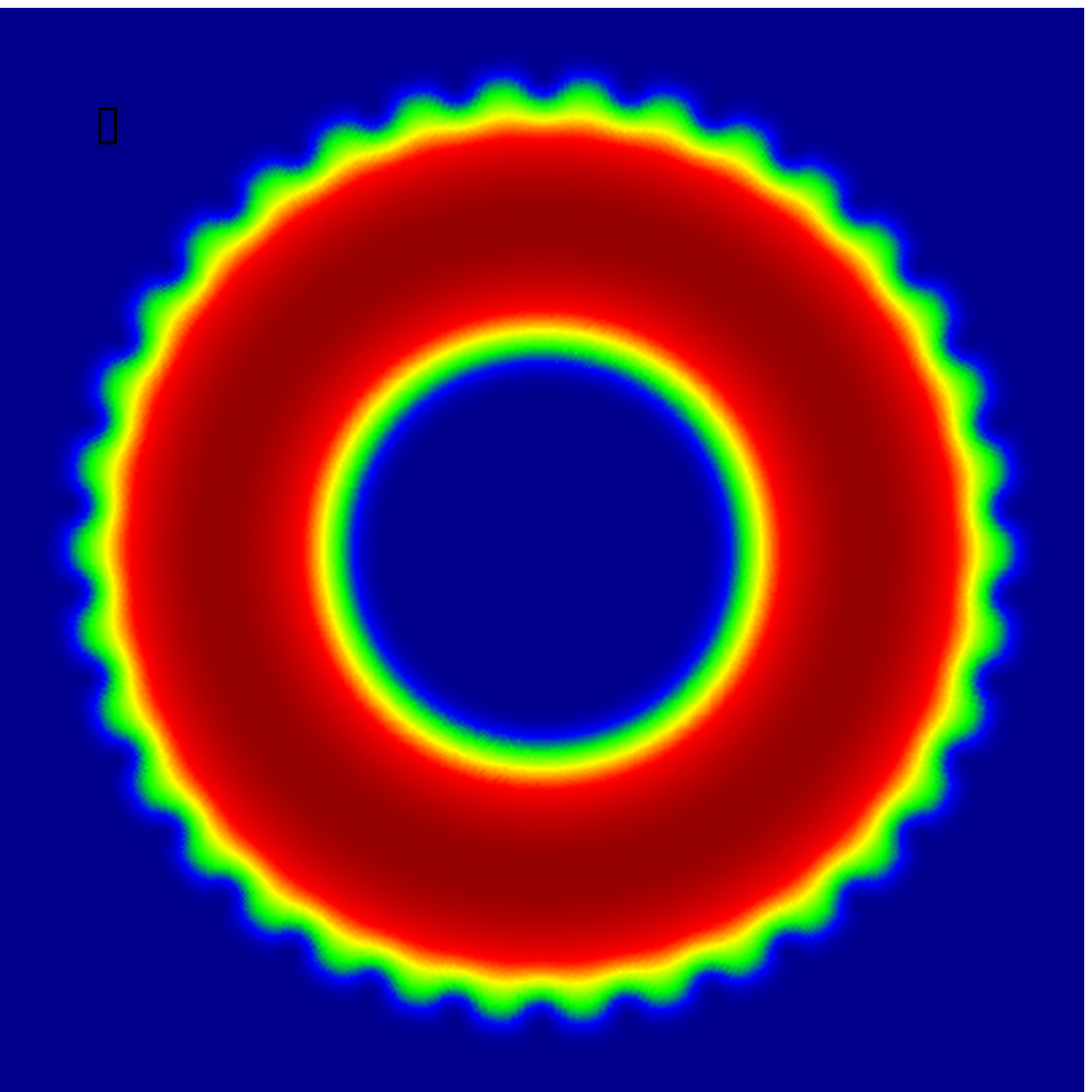}
\caption{\label{fig:spectruml}(Color online)
$(a)$ Lower branch of the excitation spectrum for $m<0$,
$r_0=12$, $g=9000$ and initial states with increasing circulation,
$\ell=0,6,11,12,13,15$.
Inset: full spectrum for $\ell=11$ (black solid line) and
surface mode model relation dispersion for outer mode (red dashed
line) or inner mode (blue dotted line).
$(b)$ Radial density profiles of the condensate (black solid
line), sound-like mode ($\ell=0$, $m=-1$, green dash-dotted line,
scaled $\times3$ for clarity), inner edge surface mode ($\ell=15$,
$m=-14$, blue dotted line) and  outer edge surface mode ($\ell=0$,
$m=-34$, red dashed line).
$(c$-$d)$ Corresponding plot of the atomic density with a population of 10\% in the inner mode $(c)$ and the outer mode $(d)$. The size of both images is $40\times 40$ in units of $a_r$. At the periphery of the gas, 14 vortices appear in $(c)$ and 34 anti-vortices in $(d)$.
}
\end{figure}

Figure~\ref{fig:spectrum} shows a typical spectrum obtained from
Eq.~\eqref{eqn:gpe2D} for a non-circulating initial state ($\ell=0$).
At small $m$ values, the lowest branch is linear,
$\omega_0(m)=m\Omega_s$, and can be associated to rotating soundlike
waves with angular velocity $\Omega_s$.
At larger $m$ values, this branch exhibits a small negative curvature
that makes the critical angular velocity smaller than the angular
speed of sound [$\Omega_c(0)<\Omega_s$].
This is not surprising as it is already the case in a linear geometry
for an inhomogeneous gas~\cite{Stringari1998,Fedichev2001}.
As $m$ increases, the radial profile of the associated density
perturbation $\delta\rho_m^0(r)$
of the lowest energy mode, where
$\delta\rho_m^\ell(r)=2\,\re{\psi_\ell(r)^*\left(u_m^\ell(r)+v_m^{\ell}(r)^*\right)}$,
is more and more localized on the outer radius 
[Fig.~\ref{fig:spectruml}(b)].
We thus expect that the mode corresponding to the critical angular
velocity will be correctly described by a surface mode model.

Following the approach of Ref.~\cite{Anglin2001}, we find the critical
velocity for a family of modes lying on the edge of a condensate.
Locally, the surface can be considered as a plane, and a surface
excitation with wave vector $k$ parallel to this plane gives a
critical linear velocity $v_c\simeq\sqrt{2}\mu^{1/6}$ in our
dimensionless units, for the critical wave vector
$k_c\simeq0.89\times\sqrt{2}\mu^{1/6}$~\cite{Anglin2001}.
In our ring-shaped geometry we identify the critical angular velocity
as $\Omega_c=v_c/r_e$, where $r_e$ is the radius at which the
excitation is localized.
Within the surface mode model, the critical angular velocity is then:
\begin{equation}
\Omega_c=\frac{\sqrt{2}\mu^{1/6}}{r_e}
\label{eqn:modelcritical}
\end{equation}
and corresponds to a critical excitation $m_c=k_c r_e$, where
$r_e\simeq r_0+R$ (respectively $r_0-R$) for an excitation lying on the outer
(respectively inner) edge of the condensate.
The dispersion relations of the inner and outer edge surface modes are
plotted as solid lines in Fig.~\ref{fig:spectrum}.
For an initial state $\ell=0$, the inner mode belongs to a higher
branch and thus does not determine the critical velocity.

\begin{figure}[t]
\includegraphics[width=\columnwidth]{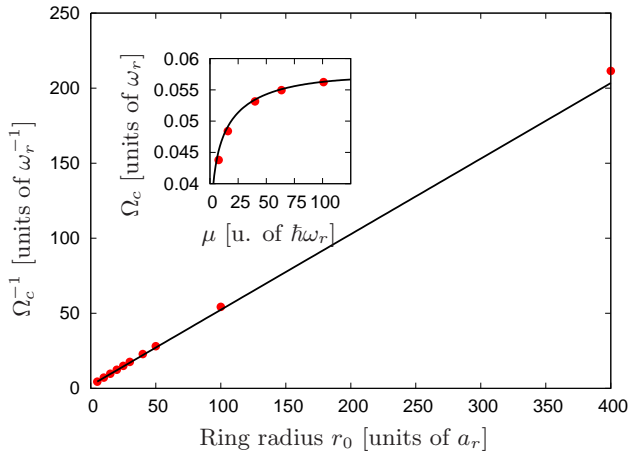}%
\caption{\label{fig:model_r0}(Color online)
Inverse of the critical angular velocity as a function of the ring
radius $r_0$, at fixed chemical potential~\cite{Note3}, as obtained
from the full numerical calculation (dots).
Inset: critical angular velocity as a function of $\mu$ for $r_0=40$.
In both graphs, the solid line is the model of
Eq.~\eqref{eqn:modelcritical} with $r_e=r_0+R$.
}
\end{figure}
\footnotetext{More precisely we keep the ratio $g/r_0$ constant so
that the chemical potential and hence the transverse profile of the
condensate are roughly constant. Indeed in the Thomas-Fermi limit
$\mu$ is expected to scale as $(g/r_0)^{2/3}$~\cite{Morizot2006}.}
Figure~\ref{fig:model_r0} shows a comparison of the full numerical
calculation of the critical velocity with the surface mode model.
The model of Eq.~\eqref{eqn:modelcritical} is in good agreement with
the numerical calculation and can thus be used to get an estimation of
the critical velocity.
We note that the agreement with the numerical calculation is better
for larger $\mu$, as shown in the inset of Fig.~\ref{fig:model_r0},
since the excitation is sufficiently localized on the
surface~\cite{Anglin2001}.

We now turn to the more complex situation of an initial state with
given circulation $\ell$, as obtained experimentally in recent
experiments~\cite{Ramanathan2011,Moulder2012}.
Figure~\ref{fig:spectruml}(a) shows the lower branch of the excitation
spectrum for initial states with increasing circulation.
As expected, for $\ell>0$, the spectrum becomes asymmetric,
excitations with $m<0$ propagating against the superflow having lower
energies than those with $m>0$, [Fig.~\ref{fig:spectruml}(a), inset].
One striking feature of this spectrum is that for a sufficiently large
$\ell$, the critical velocity is associated with a branch that crosses
the outer edge surface mode branch.
The radial profile $\delta\rho_\ell^m(r)$ shows that these modes are
located at the inner edge of the ring [see
Fig.~\ref{fig:spectruml}(b)].
Hence, depending on the initial circulation, the most probable
mechanism for dissipation implies either outer or inner modes.
Interestingly, the phase profile of the perturbation displays
antivortex patterns on the outer edge for outer modes and vortex
patterns on the inner edge for inner modes.
This supports the idea that these modes are precursors of (anti)
vortex nucleation. The total density $|\psi|^2$ with a
fraction of $10\%$ in the excited mode is displayed in
Fig.~\ref{fig:spectruml}(c) for the inner mode, and
Fig.~\ref{fig:spectruml}(d) for the outer mode. In both cases, vortices (antivortices) are located on the inner (outer) edge of the gas where the density vanishes.

This result can be understood by extending the surface mode model to
an initial state with circulation.
It is important to remark that due to the superfluid nature of the
condensate flow, the local velocity at the inner edge is larger than
the one at the outer edge.
In the frame corotating with one of these edges, where the condensate
surface is at rest, the result of the surface mode model still holds.
In the laboratory frame, the critical angular velocity is then shifted
by the angular velocity of the corotating frame:
$\Omega_r=\ell/r_e^2$, which depends on the edge considered, and may
be written
\begin{equation}
\Omega_c(\ell)=\frac{\sqrt{2}\mu^{1/6}}{r_e}-\frac{\ell}{r_e^2}.
\label{eqn:modelcrit}
\end{equation}
The result of Eq.~\eqref{eqn:modelcrit} is the sum of two
contributions: the first one arises from the surface mode model,
whereas the second one is due to the superfluid rotation of the
condensate itself.
These two terms depend with different power laws on the excited mode
radius $r_e$ and this feature explains the transition between inner
and outer edge surface excitations observed in
Fig.~\ref{fig:spectruml}(a) (see inset).

\begin{figure}[t]
\includegraphics[width=\columnwidth]{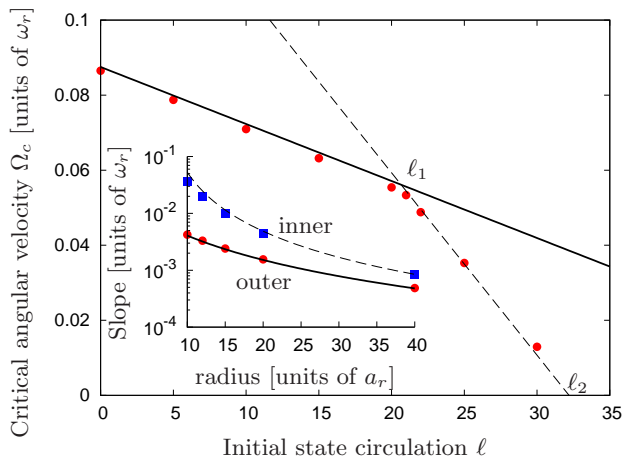}%
\caption{\label{fig:critrot}(Color online)
Critical angular velocity versus initial state circulation, as
obtained from the numerical solution (dots), for $r_0=20$ and
$g=15\,000$.
Inset: slope of the critical velocity for inner (blue squares) and
outer (red dots) modes versus the ring radius, at fixed chemical
potential~\cite{Note3}.
For both graphs, the lines are the predictions of
Eq.~\eqref{eqn:modelcrit}, for $r_e=r_0+R$ (solid line) and
$r_e=r_0-R$ (dashed line).
}
\end{figure}
Figure~\ref{fig:critrot} shows the critical angular velocity as a
function of the initial state circulation $\ell$.
The curves exhibit a piecewise linear dependence on the circulation.
The two slopes correspond, respectively, to modes lying on the outer or
inner edge.
The critical angular velocity and these slopes are compared to the
surface mode model.
The agreement is good except for small radii where the centrifugal
term $1/r$ in the Laplacian plays an important role, especially in the
case of the inner mode.
At large values of $r_0$ the difference between inner and outer modes
becomes less pronounced as the system resembles more and more an
infinite tube, where these modes are degenerate.

To interpret the data we define two thresholds.
The first threshold $\ell_1=\mu^{1/6}(r_0^2-R^2)/(\sqrt{2}r_0)$ is
obtained when Eq.~\eqref{eqn:modelcrit} gives the same value for
$r_e=r_0\pm R$. It corresponds to the frontier between the domains
where either the inner or the outer modes govern dissipation.
We find that above a second threshold
$\ell_2=\sqrt{2}\mu^{1/6}(r_0-R)$, for which the critical angular
velocity vanishes, the system is unstable, in the sense that the
computed spectra contain negative energies.
Any static perturbation of the system thus triggers dissipation and
circulation becomes highly unstable. It is important to point out that
the value of $\ell_2$ is related to the velocity of the surface mode,
not to the speed of sound.

From a practical point of view, this sets an upper limit on the
circulation that can be imprinted for a given ring radius and chemical
potential.
Therefore, even if the ring geometry is adapted to study the
persistent flow of a superfluid, it can bear only a limited amount of
circulation.

Within this model we can make predictions for the superflow stability
in recent experiments.
In the presence of static defects, the superflow will be unstable for
$\ell>\ell_2$.  
We compute this maximum allowed circulation for a stable flow with the
experimental parameters of Ref.~\cite{Ramanathan2011}, namely, $r_0=10$
and $\mu=10.9$, and find $\ell_2\simeq11$.
We may wonder if our 2D model applies to three-dimensional (3D) experiments.
The surface mode model depends only on the force at the
surface~\cite{Anglin2001}.
For a harmonically trapped BEC, this force is fully determined by the
Thomas-Fermi radius and the oscillation frequency.
A 2D model is thus expected to compare well to a 3D experiment with
the same Thomas-Fermi radius (or chemical potential).

Our results enlighten the recent work of Ref.~\cite{Piazza2009} where
a dynamical simulation of a circulating annular Bose-Einstein
condensate in the presence of a static weak link shows a dissipation
mechanism based on two critical barrier heights, associated to a
vortex-antivortex annihilation.
Our work shows that indeed a static defect can induce dissipation by
first coupling to an inner edge surface mode, allowing a vortex to
nucleate, as shown in Fig.~\ref{fig:spectruml}(c).
The excitation of the outer edge surface mode, implying
anti-vortices located all around the outer edge [see Fig.~\ref{fig:spectruml}(d)], requires a stronger
excitation.

In conclusion we have computed the Bogoliubov spectrum of circulating
annular Bose gases and obtained analytical expressions for the
critical angular velocity based on a surface mode model.
We have pointed out the role of inner and outer modes in the
determination of the critical angular velocity.
We have discussed the implications of these results to explain the
dissipation of a persistent flow and have shown that the circulation
is unstable above a given threshold.
In fact, the Landau argument~\cite{Landau1941} is quite subtle in a
ring geometry.
Indeed, as shown in this paper, we find different results for the
motion of a defect in a superfluid at rest or for a circulating
superfluid flowing through a static defect.
In the former case outer edge surface modes are always excited first,
while in the latter case the inner edge surface modes dominate.
This suggests that in an annular geometry the notion of local speed of
sound, associated with modes centered at the peak density, is not the
most relevant to discuss superfluidity.

Further work will include a numerical study of the dynamics of an
annular Bose gas in the presence of static and rotating defects to further
investigate the critical velocity.
In particular, we expect that engineering the shape of a perturbation
may help to selectively excite only the critical mode and thus more
precisely control the system.
An interesting point would be to examine the possibility of inducing a
circulation by selectively exciting this mode. Symmetrically,
while it is clear from our results that dissipation of a flow with an
initial circulation $\ell>\ell_2$ first occurs through the appearance
of vortices at the inner edge, the subsequent evolution is an
interesting open question which will require the simulation of the
full dynamics.

\begin{acknowledgments}
Laboratoire de physique des lasers is UMR 7538 of CNRS and Paris 13
University. LPL is a member of the Institut Francilien de Recherche
sur les Atomes Froids (IFRAF). H. P. acknowledges support from the
iXCore Foundation for Research.
\end{acknowledgments}


\begin{thebibliography}{3}%
\makeatletter
\providecommand \@ifxundefined [1]{%
 \@ifx{#1\undefined}
}%
\providecommand \@ifnum [1]{%
 \ifnum #1\expandafter \@firstoftwo
 \else \expandafter \@secondoftwo
 \fi
}%
\providecommand \@ifx [1]{%
 \ifx #1\expandafter \@firstoftwo
 \else \expandafter \@secondoftwo
 \fi
}%
\providecommand \natexlab [1]{#1}%
\providecommand \enquote  [1]{``#1''}%
\providecommand \bibnamefont  [1]{#1}%
\providecommand \bibfnamefont [1]{#1}%
\providecommand \citenamefont [1]{#1}%
\providecommand \href@noop [0]{\@secondoftwo}%
\providecommand \href [0]{\begingroup \@sanitize@url \@href}%
\providecommand \@href[1]{\@@startlink{#1}\@@href}%
\providecommand \@@href[1]{\endgroup#1\@@endlink}%
\providecommand \@sanitize@url [0]{\catcode `\\12\catcode `\$12\catcode
  `\&12\catcode `\#12\catcode `\^12\catcode `\_12\catcode `\%12\relax}%
\providecommand \@@startlink[1]{}%
\providecommand \@@endlink[0]{}%
\providecommand \url  [0]{\begingroup\@sanitize@url \@url }%
\providecommand \@url [1]{\endgroup\@href {#1}{\urlprefix }}%
\providecommand \urlprefix  [0]{URL }%
\providecommand \Eprint [0]{\href }%
\providecommand \doibase [0]{http://dx.doi.org/}%
\providecommand \selectlanguage [0]{\@gobble}%
\providecommand \bibinfo  [0]{\@secondoftwo}%
\providecommand \bibfield  [0]{\@secondoftwo}%
\providecommand \translation [1]{[#1]}%
\providecommand \BibitemOpen [0]{}%
\providecommand \bibitemStop [0]{}%
\providecommand \bibitemNoStop [0]{.\EOS\space}%
\providecommand \EOS [0]{\spacefactor3000\relax}%
\providecommand \BibitemShut  [1]{\csname bibitem#1\endcsname}%
\let\auto@bib@innerbib\@empty
\bibitem [{Note1()}]{Note1}%
  \BibitemOpen
  \bibinfo {note} {$g=N\protect \sqrt {8\pi }\protect \frac {a}{a_z}$ where $N$
  is the atom number, $a$ is the scattering length and $a_z$ the ground state
  size along the strongly confined direction $z$, see for example~\cite
  {Petrov2000}.}\BibitemShut {Stop}%
\bibitem [{Note2()}]{Note2}%
  \BibitemOpen
  \bibinfo {note} {The initial stationary state of circulation $\ell $ is found
  as the result of an imaginary time propagation of a test Thomas-Fermi profile
  stopped when the relative variation of the chemical potential reaches
  $10^{-12}$. The Bogoliubov-de Gennes equations are diagonalized using a C++
  implementation of the LAPACK library~\cite {Anderson1999}.}\BibitemShut
  {Stop}%
\bibitem [{Note3()}]{Note3}%
  \BibitemOpen
  \bibinfo {note} {More precisely we keep the ratio $g/r_0$ constant so that
  the chemical potential and hence the transverse profile of the condensate are
  roughly constant. Indeed in the Thomas-Fermi limit $\mu $ is expected to
  scale as $(g/r_0)^{2/3}$~\cite {Morizot2006}.}\BibitemShut {Stop}%
\end{thebibliography}%


\begin{thebibliography}{10}%
\makeatletter
\providecommand \@ifxundefined [1]{%
 \ifx #1\undefined \expandafter \@firstoftwo
 \else \expandafter \@secondoftwo
\fi
}%
\providecommand \@ifnum [1]{%
 \ifnum #1\expandafter \@firstoftwo
 \else \expandafter \@secondoftwo
\fi
}%
\providecommand \enquote [1]{``#1''}%
\providecommand \bibnamefont  [1]{#1}%
\providecommand \bibfnamefont [1]{#1}%
\providecommand \citenamefont [1]{#1}%
\providecommand\href[0]{\@sanitize\@href}%
\providecommand\@href[1]{\endgroup\@@startlink{#1}\endgroup\@@href}%
\providecommand\@@href[1]{#1\@@endlink}%
\providecommand \@sanitize [0]{\begingroup\catcode`\&12\catcode`\#12\relax}%
\@ifxundefined \pdfoutput {\@firstoftwo}{%
 \@ifnum{\z@=\pdfoutput}{\@firstoftwo}{\@secondoftwo}%
}{%
 \providecommand\@@startlink[1]{\leavevmode}%
 \providecommand\@@endlink[0]{}%
}{%
 \providecommand\@@startlink[1]{%
  \leavevmode
  \pdfstartlink
   attr{/Border[0 0 1 ]/H/I/C[0 1 1]}%
   user{/Subtype/Link/A<</Type/Action/S/URI/URI(#1)>>}%
  \relax
 }%
 \providecommand\@@endlink[0]{\pdfendlink}%
}%
\providecommand \url  [0]{\begingroup\@sanitize \@url }%
\providecommand \@url [1]{\endgroup\@href {#1}{\urlprefix}}%
\providecommand \urlprefix [0]{URL }%
\providecommand \Eprint[0]{\href }%
\@ifxundefined \urlstyle {%
  \providecommand \doi [1]{doi:\discretionary{}{}{}#1}%
}{%
  \providecommand \doi [0]{doi:\discretionary{}{}{}\begingroup
  \urlstyle{rm}\Url }%
}%
\providecommand \doibase [0]{http://dx.doi.org/}%
\providecommand \Doi[1]{\href{\doibase#1}}%
\providecommand \bibAnnote [3]{%
  \BibitemShut{#1}%
  \begin{quotation}\noindent
    \textsc{Key:}\ #2\\\textsc{Annotation:}\ #3%
  \end{quotation}%
}%
\providecommand \bibAnnoteFile [2]{%
  \IfFileExists{#2}{\bibAnnote {#1} {#2} {\input{#2}}}{}%
}%
\providecommand \typeout [0]{\immediate \write \m@ne }%
\providecommand \selectlanguage [0]{\@gobble}%
\providecommand \bibinfo [0]{\@secondoftwo}%
\providecommand \bibfield [0]{\@secondoftwo}%
\providecommand \translation [1]{[#1]}%
\providecommand \BibitemOpen[0]{}%
\providecommand \bibitemStop [0]{}%
\providecommand \bibitemNoStop [0]{.\EOS\space}%
\providecommand \EOS [0]{\spacefactor3000\relax}%
\providecommand \BibitemShut [1]{\csname bibitem#1\endcsname}%
\bibitem{Hall1957}%
  \BibitemOpen
  \bibfield{author}{%
  \bibinfo {author} {\bibfnamefont{H.~E.}\ \bibnamefont{Hall}},\ }%
  \bibfield{journal}{%
  \Doi{10.1098/rsta.1957.0024}{\bibinfo {journal} {Phil. Trans. R. Soc. A}}\ }%
  \textbf{\bibinfo {volume} {250}},\ \bibinfo {pages} {359} (\bibinfo {year}
  {1957})%
  \bibAnnoteFile{NoStop}{Hall1957}%
\bibitem{Ryu2007}%
  \BibitemOpen
  \bibfield{author}{%
  \bibinfo {author} {\bibfnamefont{C.}~\bibnamefont{Ryu}}, \bibinfo {author}
  {\bibfnamefont{M.}~\bibnamefont{Andersen}}, \bibinfo {author}
  {\bibfnamefont{P.}~\bibnamefont{Clad\'{e}}}, \bibinfo {author}
  {\bibfnamefont{V.}~\bibnamefont{Natarajan}}, \bibinfo {author}
  {\bibfnamefont{K.}~\bibnamefont{Helmerson}},\ and\ \bibinfo {author}
  {\bibfnamefont{W.}~\bibnamefont{Phillips}},\ }%
  \bibfield{journal}{%
  \Doi{10.1103/PhysRevLett.99.260401}{\bibinfo {journal} {Phys. Rev. Lett.}}\
  }%
  \textbf{\bibinfo {volume} {99}},\ \bibinfo {pages} {260401} (\bibinfo {year}
  {2007})%
  \bibAnnoteFile{NoStop}{Ryu2007}%
\bibitem{Ramanathan2011}%
  \BibitemOpen
  \bibfield{author}{%
  \bibinfo {author} {\bibfnamefont{A.}~\bibnamefont{Ramanathan}}, \bibinfo
  {author} {\bibfnamefont{K.~C.}\ \bibnamefont{Wright}}, \bibinfo {author}
  {\bibfnamefont{S.~R.}\ \bibnamefont{Muniz}}, \bibinfo {author}
  {\bibfnamefont{M.}~\bibnamefont{Zelan}}, \bibinfo {author}
  {\bibfnamefont{W.~T.}\ \bibnamefont{Hill}}, \bibinfo {author}
  {\bibfnamefont{C.~J.}\ \bibnamefont{Lobb}}, \bibinfo {author}
  {\bibfnamefont{K.}~\bibnamefont{Helmerson}}, \bibinfo {author}
  {\bibfnamefont{W.~D.}\ \bibnamefont{Phillips}},\ and\ \bibinfo {author}
  {\bibfnamefont{G.~K.}\ \bibnamefont{Campbell}},\ }%
  \bibfield{journal}{%
  \Doi{10.1103/PhysRevLett.106.130401}{\bibinfo {journal} {Phys. Rev. Lett.}}\
  }%
  \textbf{\bibinfo {volume} {106}},\ \bibinfo {pages} {130401} (\bibinfo {year}
  {2011})%
  \bibAnnoteFile{NoStop}{Ramanathan2011}%
\bibitem{Moulder2012}%
  \BibitemOpen
  \bibfield{author}{%
  \bibinfo {author} {\bibfnamefont{S.}~\bibnamefont{Moulder}}, \bibinfo
  {author} {\bibfnamefont{S.}~\bibnamefont{Beattie}}, \bibinfo {author}
  {\bibfnamefont{R.~P.}\ \bibnamefont{Smith}}, \bibinfo {author}
  {\bibfnamefont{N.}~\bibnamefont{Tammuz}},\ and\ \bibinfo {author}
  {\bibfnamefont{Z.}~\bibnamefont{Hadzibabic}},\ }%
  \bibfield{journal}{%
  \bibinfo {journal} {ArXiv preprint}}%
   (\bibinfo {year} {2012}),\
  \Eprint{http://arxiv.org/abs/1112.0334}{arXiv:1112.0334}%
  \bibAnnoteFile{NoStop}{Moulder2012}%
\bibitem{Landau1941}%
  \BibitemOpen
  \bibfield{author}{%
  \bibinfo {author} {\bibfnamefont{L.~D.}\ \bibnamefont{Landau}},\ }%
  \bibfield{journal}{%
  \bibinfo {journal} {J. Phys USSR}\ }%
  \textbf{\bibinfo {volume} {5}},\ \bibinfo {pages} {71} (\bibinfo {year}
  {1941})%
  \bibAnnoteFile{NoStop}{Landau1941}%
\bibitem{Raman1999}%
  \BibitemOpen
  \bibfield{author}{%
  \bibinfo {author} {\bibfnamefont{C.}~\bibnamefont{Raman}}, \bibinfo {author}
  {\bibfnamefont{M.}~\bibnamefont{K\"{o}hl}}, \bibinfo {author}
  {\bibfnamefont{R.}~\bibnamefont{Onofrio}}, \bibinfo {author}
  {\bibfnamefont{D.}~\bibnamefont{Durfee}}, \bibinfo {author}
  {\bibfnamefont{C.}~\bibnamefont{Kuklewicz}}, \bibinfo {author}
  {\bibfnamefont{Z.}~\bibnamefont{Hadzibabic}},\ and\ \bibinfo {author}
  {\bibfnamefont{W.}~\bibnamefont{Ketterle}},\ }%
  \bibfield{journal}{%
  \Doi{10.1103/PhysRevLett.83.2502}{\bibinfo {journal} {Phys. Rev. Lett.}}\ }%
  \textbf{\bibinfo {volume} {83}},\ \bibinfo {pages} {2502} (\bibinfo {year}
  {1999})%
  \bibAnnoteFile{NoStop}{Raman1999}%
\bibitem{Leggett1999}%
  \BibitemOpen
  \bibfield{author}{%
  \bibinfo {author} {\bibfnamefont{A.}~\bibnamefont{Leggett}},\ }%
  \bibfield{journal}{%
  \Doi{10.1103/RevModPhys.71.S318}{\bibinfo {journal} {Rev. Mod. Phys.}}\ }%
  \textbf{\bibinfo {volume} {71}},\ \bibinfo {pages} {S318} (\bibinfo {year}
  {1999})%
  \bibAnnoteFile{NoStop}{Leggett1999}%
\bibitem{Fedichev2001}%
  \BibitemOpen
  \bibfield{author}{%
  \bibinfo {author} {\bibfnamefont{P.}~\bibnamefont{Fedichev}}\ and\ \bibinfo
  {author} {\bibfnamefont{G.}~\bibnamefont{Shlyapnikov}},\ }%
  \bibfield{journal}{%
  \Doi{10.1103/PhysRevA.63.045601}{\bibinfo {journal} {Phys. Rev. A}}\ }%
  \textbf{\bibinfo {volume} {63}},\ \bibinfo {pages} {045601} (\bibinfo {year}
  {2001})%
  \bibAnnoteFile{NoStop}{Fedichev2001}%
\bibitem{Anglin2001}%
  \BibitemOpen
  \bibfield{author}{%
  \bibinfo {author} {\bibfnamefont{J.~R.}\ \bibnamefont{Anglin}},\ }%
  \bibfield{journal}{%
  \Doi{10.1103/PhysRevLett.87.240401}{\bibinfo {journal} {Phys. Rev. Lett.}}\
  }%
  \textbf{\bibinfo {volume} {87}},\ \bibinfo {pages} {240401} (\bibinfo {year}
  {2001})%
  \bibAnnoteFile{NoStop}{Anglin2001}%
\bibitem{Donnelly1966}%
  \BibitemOpen
  \bibfield{author}{%
  \bibinfo {author} {\bibfnamefont{R.}~\bibnamefont{Donnelly}}\ and\ \bibinfo
  {author} {\bibfnamefont{A.}~\bibnamefont{Fetter}},\ }%
  \bibfield{journal}{%
  \Doi{10.1103/PhysRevLett.17.747}{\bibinfo {journal} {Phys. Rev. Lett.}}\ }%
  \textbf{\bibinfo {volume} {17}},\ \bibinfo {pages} {747} (\bibinfo {year}
  {1966})%
  \bibAnnoteFile{NoStop}{Donnelly1966}%
\bibitem{Wright2000}%
  \BibitemOpen
  \bibfield{author}{%
  \bibinfo {author} {\bibfnamefont{E.}~\bibnamefont{Wright}}, \bibinfo {author}
  {\bibfnamefont{J.}~\bibnamefont{Arlt}},\ and\ \bibinfo {author}
  {\bibfnamefont{K.}~\bibnamefont{Dholakia}},\ }%
  \bibfield{journal}{%
  \Doi{10.1103/PhysRevA.63.013608}{\bibinfo {journal} {Phys. Rev. A}}\ }%
  \textbf{\bibinfo {volume} {63}},\ \bibinfo {pages} {013608} (\bibinfo {year}
  {2000})%
  \bibAnnoteFile{NoStop}{Wright2000}%
\bibitem{Morizot2006}%
  \BibitemOpen
  \bibfield{author}{%
  \bibinfo {author} {\bibfnamefont{O.}~\bibnamefont{Morizot}}, \bibinfo
  {author} {\bibfnamefont{Y.}~\bibnamefont{Colombe}}, \bibinfo {author}
  {\bibfnamefont{V.}~\bibnamefont{Lorent}}, \bibinfo {author}
  {\bibfnamefont{H.}~\bibnamefont{Perrin}},\ and\ \bibinfo {author}
  {\bibfnamefont{B.}~\bibnamefont{Garraway}},\ }%
  \bibfield{journal}{%
  \Doi{10.1103/PhysRevA.74.023617}{\bibinfo {journal} {Phys. Rev. A}}\ }%
  \textbf{\bibinfo {volume} {74}},\ \bibinfo {pages} {023617} (\bibinfo {year}
  {2006})%
  \bibAnnoteFile{NoStop}{Morizot2006}%
\bibitem{Lesanovsky2007}%
  \BibitemOpen
  \bibfield{author}{%
  \bibinfo {author} {\bibfnamefont{I.}~\bibnamefont{Lesanovsky}}\ and\ \bibinfo
  {author} {\bibfnamefont{W.}~\bibnamefont{von Klitzing}},\ }%
  \bibfield{journal}{%
  \Doi{10.1103/PhysRevLett.99.083001}{\bibinfo {journal} {Phys. Rev. Lett.}}\
  }%
  \textbf{\bibinfo {volume} {99}},\ \bibinfo {pages} {083001} (\bibinfo {month}
  {Aug.}\ \bibinfo {year} {2007})%
  \bibAnnoteFile{NoStop}{Lesanovsky2007}%
\bibitem{Heathcote2008}%
  \BibitemOpen
  \bibfield{author}{%
  \bibinfo {author} {\bibfnamefont{W.~H.}\ \bibnamefont{Heathcote}}, \bibinfo
  {author} {\bibfnamefont{E.}~\bibnamefont{Nugent}}, \bibinfo {author}
  {\bibfnamefont{B.~T.}\ \bibnamefont{Sheard}},\ and\ \bibinfo {author}
  {\bibfnamefont{C.~J.}\ \bibnamefont{Foot}},\ }%
  \bibfield{journal}{%
  \Doi{10.1088/1367-2630/10/4/043012}{\bibinfo {journal} {New J. Phys.}}\ }%
  \textbf{\bibinfo {volume} {10}},\ \bibinfo {pages} {043012} (\bibinfo {year}
  {2008})%
  \bibAnnoteFile{NoStop}{Heathcote2008}%
\bibitem{Sherlock2011}%
  \BibitemOpen
  \bibfield{author}{%
  \bibinfo {author} {\bibfnamefont{B.~E.}\ \bibnamefont{Sherlock}}, \bibinfo
  {author} {\bibfnamefont{M.}~\bibnamefont{Gildemeister}}, \bibinfo {author}
  {\bibfnamefont{E.}~\bibnamefont{Owen}}, \bibinfo {author}
  {\bibfnamefont{E.}~\bibnamefont{Nugent}},\ and\ \bibinfo {author}
  {\bibfnamefont{C.~J.}\ \bibnamefont{Foot}},\ }%
  \bibfield{journal}{%
  \Doi{10.1103/PhysRevA.83.043408}{\bibinfo {journal} {Phys. Rev. A}}\ }%
  \textbf{\bibinfo {volume} {83}},\ \bibinfo {pages} {043408} (\bibinfo {year}
  {2011})%
  \bibAnnoteFile{NoStop}{Sherlock2011}%
\bibitem{Piazza2009}%
  \BibitemOpen
  \bibfield{author}{%
  \bibinfo {author} {\bibfnamefont{F.}~\bibnamefont{Piazza}}, \bibinfo {author}
  {\bibfnamefont{L.}~\bibnamefont{Collins}},\ and\ \bibinfo {author}
  {\bibfnamefont{A.}~\bibnamefont{Smerzi}},\ }%
  \bibfield{journal}{%
  \Doi{10.1103/PhysRevA.80.021601}{\bibinfo {journal} {Phys. Rev. A}}\ }%
  \textbf{\bibinfo {volume} {80}},\ \bibinfo {pages} {021601} (\bibinfo {year}
  {2009})%
  \bibAnnoteFile{NoStop}{Piazza2009}%
\bibitem{Schenke2011}%
  \BibitemOpen
  \bibfield{author}{%
  \bibinfo {author} {\bibfnamefont{C.}~\bibnamefont{Schenke}}, \bibinfo
  {author} {\bibfnamefont{A.}~\bibnamefont{Minguzzi}},\ and\ \bibinfo {author}
  {\bibfnamefont{F.~W.~J.}\ \bibnamefont{Hekking}},\ }%
  \bibfield{journal}{%
  \Doi{10.1103/PhysRevA.84.053636}{\bibinfo {journal} {Phys. Rev. A}}\ }%
  \textbf{\bibinfo {volume} {84}},\ \bibinfo {pages} {053636} (\bibinfo {year}
  {2011})%
  \bibAnnoteFile{NoStop}{Schenke2011}%
\bibitem{Fetter1967}%
  \BibitemOpen
  \bibfield{author}{%
  \bibinfo {author} {\bibfnamefont{A.~L.}\ \bibnamefont{Fetter}},\ }%
  \bibfield{journal}{%
  \Doi{10.1103/PhysRev.153.285}{\bibinfo {journal} {Phys. Rev.}}\ }%
  \textbf{\bibinfo {volume} {153}},\ \bibinfo {pages} {285} (\bibinfo {month}
  {Jan}\ \bibinfo {year} {1967})%
  \bibAnnoteFile{NoStop}{Fetter1967}%
\bibitem{Cozzini2006}%
  \BibitemOpen
  \bibfield{author}{%
  \bibinfo {author} {\bibfnamefont{M.}~\bibnamefont{Cozzini}}, \bibinfo
  {author} {\bibfnamefont{B.}~\bibnamefont{Jackson}},\ and\ \bibinfo {author}
  {\bibfnamefont{S.}~\bibnamefont{Stringari}},\ }%
  \bibfield{journal}{%
  \Doi{10.1103/PhysRevA.73.013603}{\bibinfo {journal} {Phys. Rev. A}}\ }%
  \textbf{\bibinfo {volume} {73}},\ \bibinfo {pages} {013603} (\bibinfo {year}
  {2006})%
  \bibAnnoteFile{NoStop}{Cozzini2006}%
\bibitem{Aftalion2010}%
  \BibitemOpen
  \bibfield{author}{%
  \bibinfo {author} {\bibfnamefont{A.}~\bibnamefont{Aftalion}}\ and\ \bibinfo
  {author} {\bibfnamefont{P.}~\bibnamefont{Mason}},\ }%
  \bibfield{journal}{%
  \Doi{10.1103/PhysRevA.81.023607}{\bibinfo {journal} {Phys. Rev. A}}\ }%
  \textbf{\bibinfo {volume} {81}},\ \bibinfo {pages} {023607} (\bibinfo {year}
  {2010})%
  \bibAnnoteFile{NoStop}{Aftalion2010}%
\bibitem{Mathey2010}%
  \BibitemOpen
  \bibfield{author}{%
  \bibinfo {author} {\bibfnamefont{L.}~\bibnamefont{Mathey}}, \bibinfo {author}
  {\bibfnamefont{A.}~\bibnamefont{Ramanathan}}, \bibinfo {author}
  {\bibfnamefont{K.~C.}\ \bibnamefont{Wright}}, \bibinfo {author}
  {\bibfnamefont{S.~R.}\ \bibnamefont{Muniz}}, \bibinfo {author}
  {\bibfnamefont{W.~D.}\ \bibnamefont{Phillips}},\ and\ \bibinfo {author}
  {\bibfnamefont{C.~W.}\ \bibnamefont{Clark}},\ }%
  \bibfield{journal}{%
  \Doi{10.1103/PhysRevA.82.033607}{\bibinfo {journal} {Phys. Rev. A}}\ }%
  \textbf{\bibinfo {volume} {82}},\ \bibinfo {pages} {033607} (\bibinfo {month}
  {Sep}\ \bibinfo {year} {2010})%
  \bibAnnoteFile{NoStop}{Mathey2010}%
\bibitem{Note1}%
  \BibitemOpen
  \bibinfo {note} {$g=N\protect \sqrt {8\pi }\protect \frac {a}{a_z}$ where $N$
  is the atom number, $a$ is the scattering length and $a_z$ the ground state
  size along the strongly confined direction $z$, see for example~\cite
  {Petrov2000}.}%
  \bibAnnoteFile{Stop}{Note1}%
\bibitem{Note2}%
  \BibitemOpen
  \bibinfo {note} {The initial stationary state of circulation $\ell $ is found
  as the result of an imaginary time propagation of a test Thomas-Fermi profile
  stopped when the relative variation of the chemical potential reaches
  $10^{-12}$. The Bogoliubov-de Gennes equations are diagonalized using a C++
  implementation of the LAPACK library~\cite {Anderson1999}.}%
  \bibAnnoteFile{Stop}{Note2}%
\bibitem{Stringari1998}%
  \BibitemOpen
  \bibfield{author}{%
  \bibinfo {author} {\bibfnamefont{S.}~\bibnamefont{Stringari}},\ }%
  \bibfield{journal}{%
  \Doi{10.1103/PhysRevA.58.2385}{\bibinfo {journal} {Phys. Rev. A}}\ }%
  \textbf{\bibinfo {volume} {58}},\ \bibinfo {pages} {2385} (\bibinfo {year}
  {1998})%
  \bibAnnoteFile{NoStop}{Stringari1998}%
\bibitem{Note3}%
  \BibitemOpen
  \bibinfo {note} {More precisely we keep the ratio $g/r_0$ constant so that
  the chemical potential and hence the transverse profile of the condensate are
  roughly constant. Indeed in the Thomas-Fermi limit $\mu $ is expected to
  scale as $(g/r_0)^{2/3}$~\cite {Morizot2006}.}%
  \bibAnnoteFile{Stop}{Note3}%
\bibitem{Petrov2000}%
  \BibitemOpen
  \bibfield{author}{%
  \bibinfo {author} {\bibfnamefont{D.~S.}\ \bibnamefont{Petrov}}, \bibinfo
  {author} {\bibfnamefont{M.}~\bibnamefont{Holzmann}},\ and\ \bibinfo {author}
  {\bibfnamefont{G.~V.}\ \bibnamefont{Shlyapnikov}},\ }%
  \bibfield{journal}{%
  \Doi{10.1103/PhysRevLett.84.2551}{\bibinfo {journal} {Phys. Rev. Lett.}}\ }%
  \textbf{\bibinfo {volume} {84}},\ \bibinfo {pages} {2551} (\bibinfo {year}
  {2000})%
  \bibAnnoteFile{NoStop}{Petrov2000}%
\bibitem{Anderson1999}%
  \BibitemOpen
  \bibfield{author}{%
  \bibinfo {author} {\bibfnamefont{E.}~\bibnamefont{Anderson}}, \bibinfo
  {author} {\bibfnamefont{Z.}~\bibnamefont{Bai}}, \bibinfo {author}
  {\bibfnamefont{C.}~\bibnamefont{Bischof}}, \bibinfo {author}
  {\bibfnamefont{S.}~\bibnamefont{Blackford}}, \bibinfo {author}
  {\bibfnamefont{J.}~\bibnamefont{Demmel}}, \bibinfo {author}
  {\bibfnamefont{J.}~\bibnamefont{Dongarra}}, \bibinfo {author}
  {\bibfnamefont{J.}~\bibnamefont{Du~Croz}}, \bibinfo {author}
  {\bibfnamefont{A.}~\bibnamefont{GreenBaum}}, \bibinfo {author}
  {\bibfnamefont{S.}~\bibnamefont{Hammarling}}, \bibinfo {author}
  {\bibfnamefont{A.}~\bibnamefont{McKenney}},\ and\ \bibinfo {author}
  {\bibfnamefont{D.}~\bibnamefont{Sorensen}},\ }%
  \emph{\bibinfo {title} {{LAPACK} Users' Guide}},\ \bibinfo {edition} {3rd}\
  ed.\ (\bibinfo {publisher} {S. I. A. M.},\ \bibinfo {address} {Philadelphia,
  PA},\ \bibinfo {year} {1999})%
  \bibAnnoteFile{NoStop}{Anderson1999}%
\end{thebibliography}
%

\end{document}